\documentclass[prb,twocolumn,showpacs,preprintnumbers,amssymb,superscriptaddress]{revtex4}
\usepackage{graphicx}
\usepackage[tbtags]{amsmath}
\usepackage{bm}

\begin{document}
\author{E. Il'ichev}
\email{ilichev@ipht-jena.de}
\affiliation{%
Institute for Physical High Technology, P.O. Box 100239, D-07702
Jena, Germany}
\title
{Flux qubit as a sensor for a magnetometer with quantum limited
sensitivity}
\author{Ya. S. Greenberg}
\affiliation{Novosibirsk State Technical University, 20 K. Marx
Ave., 630092 Novosibirsk, Russia.}

\date{\today}
\begin{abstract} We propose to use the quantum properties of a
superconducting flux qubit in the construction of a magnetometer
with quantum limited sensitivity. The main advantage of a flux
qubit is that its noise is rather low, and its transfer functions
relative to the measured flux can be made to be about
$10$mV/$\Phi_0$, which is an order of magnitude more than the best
value for  a conventional SQUID magnetometer. We analyze here the
voltage-to-flux, the phase-to-flux transfer functions and the main
noise sources. We show that the experimental characteristics of a
flux qubit, obtained in recent experiments, allow the use of a
flux qubit as magnetometer with energy resolution close to the
Planck constant.

\end{abstract}

\maketitle

Josephson-junction qubits are known to be candidates for
solid-state quantum computing circuits \cite{Makhlin}. However,
owing to their unique quantum properties these devices undoubtedly
can be used as sensitive detectors of different physical
quantities, such as quantum environmental noise \cite{Astaf} or
low frequency fluctuations of the junction critical current
\cite{Harl}. Here we propose to use a Josephson-junction flux
qubit as a sensitive detector of magnetic flux \cite{Shnyr}. We
show that the present state-of-art allows one to obtain the energy
sensitivity of such a detector in the order of the Planck
constant.

A flux qubit \cite{Orlando, Mooij, Mooij1} consists of three
Josephson junctions in a loop with very small inductance~$L$,
typically in the pH range. This ensures an effective decoupling
from the environment. Two junctions have an equal critical current
$I_\mathrm{c}$ and (effective) capacitance~$C$, while those of the
third junction are slightly smaller: $\alpha I_\mathrm{c}$ and
$\alpha C$, with $0.5<\alpha<1$.  At sufficiently low temperatures
(typically $T\approx (10\sim 30)$ mK) when $\Phi_\mathrm{x}$, the
external flux applied to the qubit loop, is in the close vicinity
of $\Phi_0/2$ ($\Phi_0=h/2e$ is the flux quantum, $h$ is the
Planck constant) the system has two low-lying quantum states $E_-$
and $E_+$. The energy gap of the flux qubits,
$\Delta/h=(E_+-E_-)/h$ is of the order of several GHz. Below we
assume $k_BT<<\Delta$ ($k_B$ is the Boltzmann constant), so that
the qubit is definitely in its ground state $E_-$ \cite{Temp}.

\begin{figure}
\includegraphics[width=7cm]{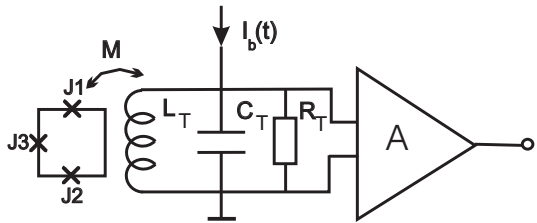}
\caption{\small Flux qubit coupled to a tank.}\label{fig1}
\end{figure}

For experimental characterization the flux qubit is inductively
coupled through a mutual inductance $M$ to an $LC$ tank circuit
with known inductance $L_\mathrm{T}$, capacitance $C_\mathrm{T}$,
and quality $Q$ (Fig.~\ref{fig1}).

The resonant characteristics of the tank circuit (frequency, phase
shift, etc.) are sensitive to the qubit inductance and therefore
to the external flux $\Phi_x$.

It was shown in \cite{Green} that the amplitude $v$ and the phase
$\chi$ of the output signal $V(t)=v\cos(\omega t+\chi)$ are
coupled by the equations:

\begin{gather}
  v^2\left(1+4Q^2\xi^2(f_\mathrm{x})\right)
    =I_0^2\omega_\mathrm{T}^2L_\mathrm{T}^2Q^2\label{V-I}\\
  \tan\chi=2Q\xi(f_\mathrm{x})\;,\label{phi-I}
\end{gather}

where $I_0$ is the amplitude of the driving current
($I_\mathrm{b}(t)=I_0\cos\omega t$), and $\omega$ and $\omega_T$
are a driving frequency and the resonant frequency of the tank
circuit, respectively.

 It is worth noting that the the scheme in Fig.
\ref{fig1} and Eqs. (\ref{V-I}), (\ref{phi-I}) are quite similar
to those for a conventional RF SQUID. The only difference is in
the expression for a flux-dependent frequency detuning $\xi(f_x)$.
This depends on the qubit parameters as \cite{Green}:

\begin{equation}\label{detuning}
  \xi(f_\mathrm{x})=\xi_0-k^2\frac{LI_\mathrm{c}^2}{\Delta}
  \left(\frac{\lambda}{2\pi}\right)^{\!2}F(f_\mathrm{x})\;,
\end{equation}
$\xi_0=(\omega_\mathrm{T}{-}\omega)/\omega_\mathrm{T}$,
$f_x=\Phi_x/\Phi_0-\frac{1}{2}$, and
\begin{equation}\label{F}
  F(f_\mathrm{x})=\frac{1}{\pi}\int_0^{2\pi}\!\!d\phi\,
  \frac{\cos^2\phi}{\bigl[1+\eta^2\left(f_\mathrm{x}+\gamma\sin\phi\right)^2\bigr]^{3/2}}\;,
\end{equation}
with $\eta=2E_\mathrm{J}\lambda/\Delta$ and $\gamma=MI_0Q/\Phi_0$.
The expression for $\lambda$, which depends on $\alpha$, $I_C$ and
$C$ is given in \cite{Green}.

Therefore, the main effect of the qubit-tank interaction is a
shift of the tank resonance. This results in a dip in the
voltage-to-flux and phase-to-flux characteristics which have been
confirmed by experiments \cite{Graj}.

.

\begin{figure}
\includegraphics[width=7 cm]{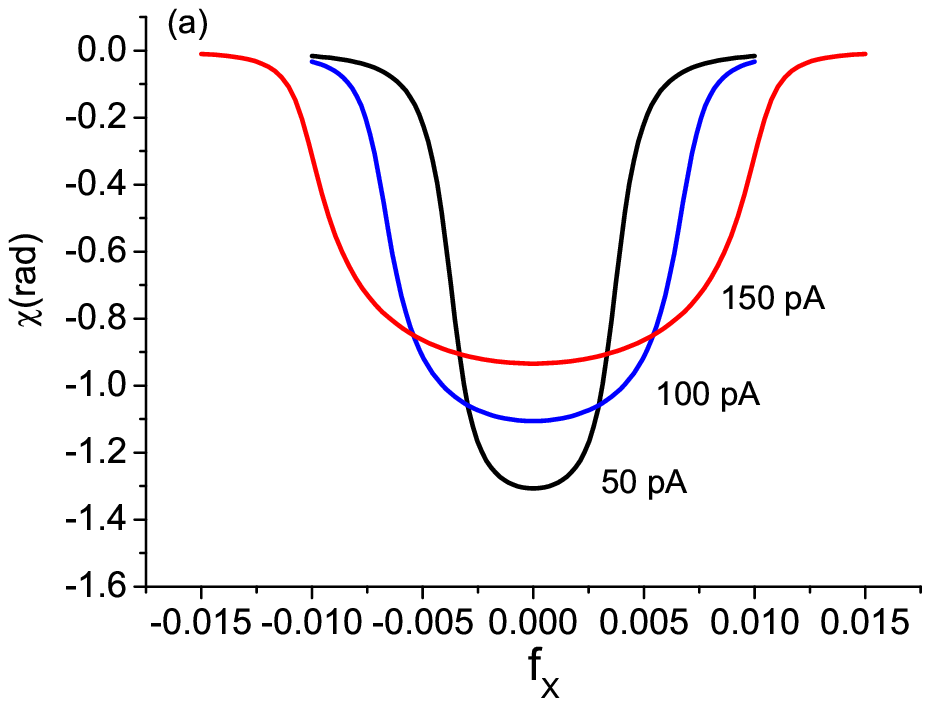}
\includegraphics[width=7 cm]{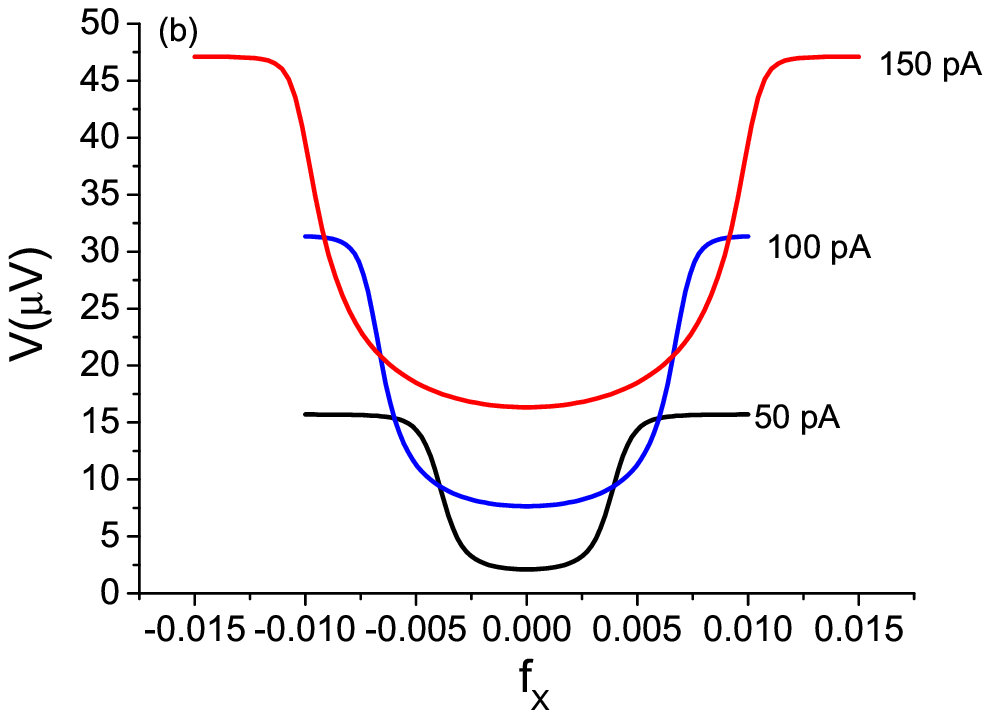}
\caption{\label{fig2} (Color online).  Tank phase $\chi$ (a) and
tank voltage $V$ (b) vs bias flux $f_x$ for three values of bias
current $I_0$. The gap frequency $\Delta/h=2$GHz.}
\end{figure}

Theoretical phase-to-flux $\chi(f_x)$ (PFC) and voltage-to-flux
(VFC) $v(f_x)$ dependencies at resonance
$\omega=\omega_\mathrm{T}$, are shown in Fig.~\ref{fig2} for three
values of the amplitude of the bias current $I_0$. The graphs
hasve been calculated from Eqs.(\ref{V-I}), (\ref{phi-I}) for the
following qubit-tank parameters: $I_c=400$ nA, $\alpha=0.8$,
$L=40$ pH, $L_T=50$~nH, $Q=10000$, $\omega_T/2\pi=100$ MHz,
$\Delta/h=2$GHz, and $k=10^{-2}$.

\begin{figure}
\includegraphics[width=7 cm]{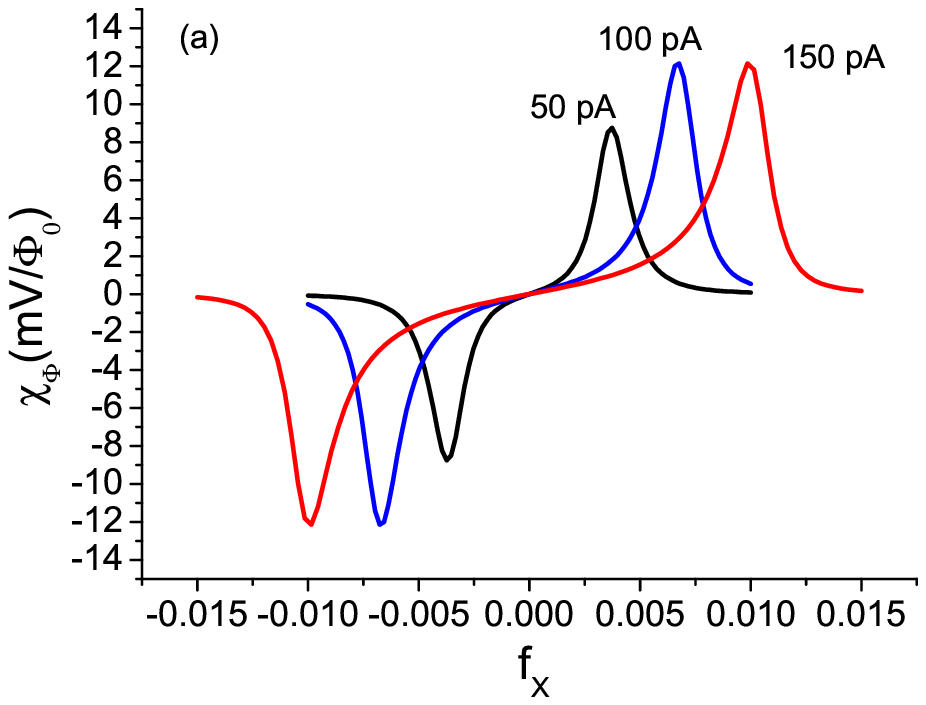}
\includegraphics[width=7 cm]{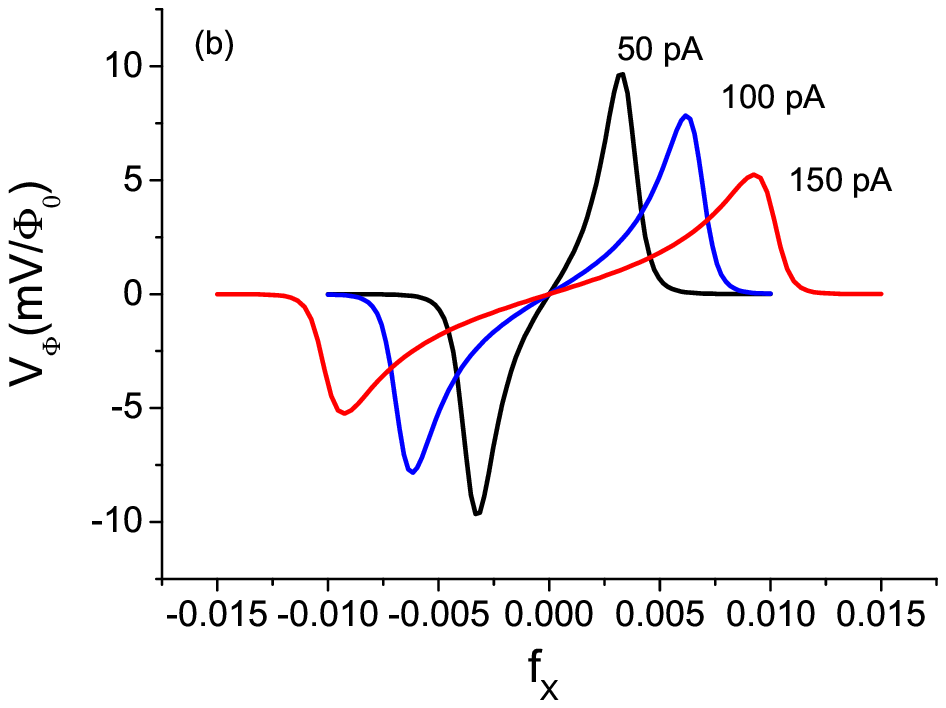}
\caption{\label{fig3} (Color online).  Phase-to-flux $\chi_{\Phi}$
(a) and voltage-to-flux $V_{\Phi}$ (b) transfer functions for
three values of bias current $I_0$. The gap frequency
$\Delta/h=2$GHz.}
\end{figure}

The advantage of a qubit magnetometer over a conventional SQUID
magnetometer is in the very steep dependence of its VFC and PFC.
In the flux locked loop operation of a magnetometer the working
point is set at a fixed value of $\Phi_X$ where the slope of VFC
or PFC is maximum. The output signal $\delta V$ is proportional to
the change $\delta\Phi_X$ of the measured flux. In principle two
modes of detection are possible: voltage mode, where $\delta
V=V_\Phi \delta\Phi_X$, and the phase mode, where $\delta
V=\chi_\Phi \delta\Phi_X$. The qubit transfer functions
$\chi_\Phi=v\partial \chi/\partial \Phi_X$ and $V_\Phi=\partial
v/\partial \Phi_X$  are shown in Fig. \ref{fig3} for the same
qubit-tank parameters as those used in Fig. \ref{fig2}. It is seen
that qubit transfer functions can exceed $10$ mV$/\Phi_0$. This
value should be compared with $1$ mV$/\Phi_0$, the best value
obtained for a DC SQUID with additional positive feedback
\cite{Drung}.

The flux and energy sensitivity depend on the main noise sources,
which come from the low frequency fluctuations of the junction
critical current, $S_{I_C}$, and from the voltage noise, $S_V$ and
the current noise, $S_I$ of the preamplifier, where $S_{I_C}$,
$S_V$ and $S_I$ are the corresponding spectral densities. The
fluctuations of $I_C$ result in the fluctuating flux in the qubit
loop $S_{\Phi,I_C}^{1/2}=LS_{I_C}^{1/2}$. For $I_C=400$ nA,
junction area $\emph{A}=0.12\mu$m$^2$, $T=0.1$ K we estimate for
three-junction flux qubit (see Eq.18 in \cite{Harl})
$S_{\Phi,I_C}^{1/2}\approx 2\times 10^{-8}\Phi_0/$Hz$^{1/2}$ at 1
Hz. We will see that this value is almost one order of magnitude
smaller than the noise from a preamplifier. Therefore, the self
noise of the qubit can be neglected. The contribution of the
voltage noise of the preamplifier to the  flux resolution referred
to the input is $S_\Phi^V=S_V/V_\Phi^2$ or
$S_\Phi^V=S_V/\chi_\Phi^2$ depending on the detection mode. The
current noise of preamplifier which is related to its noise
temperature $T_N$, $S_I=4k_BT_N/R_T$, ($R_T=\omega_TL_TQ$),
contributes via two mechanisms. The first one comes from magnetic
coupling between the tank inductance and the inductance of the
qubit loop $S_\Phi^I=M^2Q^2S_I$. This contribution cannot be
separated from the measured flux. The second mechanism contributes
through a voltage noise induced by the current noise of the
preamplifier across the dynamic resistance of the tank
$S_\Phi^D=R_D^2S_I/V_\Phi^2$ (or $S_\Phi^D=R_D^2S_I/\chi_\Phi^2$),
where $R_D=\partial v/\partial I_0$. By combining these three
mechanisms we obtain for the flux sensitivity:
\begin{equation}\label{sens}
    S_\Phi=M^2Q^2S_I+S_V/V_\Phi^2+R_D^2S_I/V_\Phi^2
\end{equation}
where $R_D$ is approximately equal to $R_T$, the resistance of
unloaded tank. In the case of the phase mode detection we should
substitute in (\ref{sens}) $\chi_\Phi$ for $V_\Phi$.

\begin{figure}
  \includegraphics[width=7 cm]{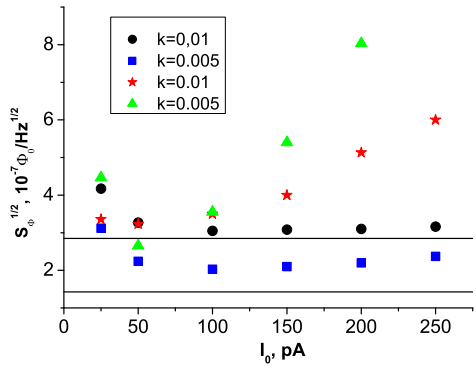}\\
  \caption{(Color online). Flux sensitivity of qubit magnetometer.
  Phase detection mode is shown by boxes and circles.
  The stars and triangles are for the voltage detection.
  Two straight lines show the level of $S_{\Phi,1}$ for $k=0.01$
  (upper line) and $k=0.005$ (lower line).}\label{Fig6}
\end{figure}

For the estimation we take $S_V^{1/2}=0.2$ nV/Hz$^{1/2}$, and
$T_N=0.1 $K \cite{Oukh}, with the other parameters being the same
as for Figs. \ref{fig2}, \ref{fig3}. The inspection of Eq.
(\ref{sens}) shows that the main contribution to the flux noise
comes from the first term:
\begin{equation}\label{fn}
    S_{\Phi,1}=M^2Q^2S_I=k^2Q\frac{4k_BT_NL}{\omega_T}
\end{equation}
This contribution does not depend on the position of the working
point and for our parameters it gives $S_{\Phi,1}^{1/2}=2.8\times
10^{-7} \Phi_0/$Hz$^{1/2}$. The influence of the last two terms in
Eq. (\ref{sens}) depends on the position of the working point and
on the bias current amplitude $I_0$. In general, the contribution
of these terms is nonnegligible. The  total flux noise dependence
on the amplitude of bias current is shown in Fig. \ref{Fig6}.
As seen from the figure phase detection in general is more
favorable than voltage detection. It gives lower noise and is
weakly sensitive to $I_0$. The flux resolution can be improved by
a decrease of $S_{\Phi,1}$ (see Eq. (\ref{fn})) upon optimization
of $k$, $Q$, $\omega_T$, or $L$. However, it does not necessarily
lead to a decrease of the total noise, since the transfer
functions also depend on these parameters. (See, for example, the
$k=005$ curve  in Fig. \ref{Fig6} for voltage detection). An
increase of the bias frequency $\omega_T$ can also give an
improved flux resolution. However, for the qubit to remain in the
ground state the condition $\omega_T<<\Delta/h$ should hold. We
also made calculations for $\omega_T=200$MHz with k=0.01, Q=1000,
with other parameters being unchanged. We obtain at $I_0=200$pA
for the phase detection $S_{\Phi}^{1/2}=1.6\times 10^{-7}
\Phi_0/$Hz$^{1/2}$, which for $L=40$pH corresponds to the energy
sensitivity $\varepsilon=S_{\Phi}/2L=1.3\times10^{-33}$J/Hz=$2h$.
These values should be compared with those for conventional
SQUIDs: $S_{\Phi}^{1/2}\approx 10^{-6} \Phi_0/$Hz$^{1/2}$,
$\varepsilon\approx 10^{-32}$J/Hz\cite{handbook}.

In summary, we have shown that a superconducting flux qubit can be
developed as a sensor of magnetic flux with an energy sensitivity
close to the Planck constant.

We thank A. Izmalkov, M. Grajcar and D. Drung for fruitful
discussions, and V. I. Shnyrkov for providing us with the
manuscript of his paper prior to publication. E.I. thanks the EU
for support through the RSFQubit and EuroSQIP projects. Ya. S. G.
acknowledges the financial support from the ESF under grant No.
1030.


\begin{thebibliography}{99}
\bibitem{Temp} In principle, this condition can be sufficiently
relaxed. In Ref. 9. it was shown that when the temperature was
increased from $T=10$ mK to $T=200$ mK  the width of the phase dip
and therefore the steepness of the phase curve remained
unchanched.
\end{thebibliography}
\end{document}